\documentclass[aps, prl, twocolumn, floatfix, nofootinbib, superscriptaddress]{revtex4-1}

\usepackage{amsmath,amsfonts,amssymb,bm}
\usepackage{dsfont}
\usepackage{graphicx}
\usepackage{color}
\usepackage{soul}

\usepackage[mathscr,scaled=1.15]{urwchancal}
\DeclareFontFamily{OT1}{pzc}{}
\DeclareFontShape{OT1}{pzc}{m}{it}%
{<-> s * [1.15] pzcmi7t}{}
\DeclareMathAlphabet{\mathpzc}{OT1}{pzc}{m}{it}

\definecolor{purple}{rgb}{0.5,0,0.5}
\definecolor{blue}{rgb}{0.0,0,0.9}
\definecolor{prdblue}{rgb}{0.133,0.118,0.498}
\usepackage[colorlinks=true, pdfstartview=FitV, linkcolor=prdblue, citecolor= prdblue, urlcolor=prdblue]{hyperref}

\begin{document}

\title{Parton distribution amplitudes: revealing diquarks in the proton and Roper resonance}

\author{C\'edric Mezrag}
\email[]{cedric.mezrag@roma1.infn.it}
\affiliation{
Istituto Nazionale di Fisica Nucleare, Sezione di Roma, P.\,le A.\,Moro 2, I-00185 Roma, Italy}
\affiliation{Physics Division, Argonne National Laboratory, Argonne, Illinois
60439, USA}

\author{Jorge Segovia}
\email[]{jsegovia@ifae.es}
\affiliation{
Institut de F\'{\i}sica d'Altes Energies (IFAE) and Barcelona Institute of Science and Technology (BIST),
Universitat Aut\`onoma de Barcelona, E-08193 Bellaterra (Barcelona), Spain
}

\author{Lei Chang}
\email[]{leichang@nankai.edu.cn}
\affiliation{School of Physics, Nankai University, Tianjin 300071, China}

\author{Craig D. Roberts}
\email[]{cdroberts@anl.gov}
\affiliation{Physics Division, Argonne National Laboratory, Argonne, Illinois
60439, USA}

\date{22 November 2017}

\begin{abstract}
We present the first quantum field theory calculation of the pointwise behaviour of the leading-twist parton distribution amplitudes (PDAs) of the proton and its lightest radial excitation.  The proton's PDA is a broad, concave function, whose maximum is shifted relative to the peak in QCD's conformal limit expression for this PDA; an effect which signals the presence of \emph{both} scalar and pseudovector diquark correlations in the nucleon, with the scalar generating around 60\% of the proton's normalisation.  The radial-excitation is constituted similarly, and the pointwise form of its PDA, which is negative on a material domain, is the result of marked interferences between the contributions from both types of diquark; particularly, the locus of zeros that highlights its character as a radial excitation.  These features originate with the emergent phenomenon of dynamical chiral-symmetry breaking in the Standard Model.
\end{abstract}



\maketitle

\noindent\emph{1.\,Introduction}\,---\,Wave functions provide insights into composite systems, \emph{e.g}.\ they express the presence and extent of correlations between constituents, and their signature in scattering processes; and thereby bridge experiment and theory, delivering understanding from what might otherwise seem arcane observations.  This is true within quantum chromodynamics (QCD), the quantum field theory describing strong interactions; but there are difficulties.  Everyday hadrons ($p\!=\,$proton, neutron, \emph{etc}.) are constituted from up ($u$) and down ($d$) valence-quarks; but the Higgs boson generates current-masses for these fermions which are more than 100-times smaller than the scale associated with the composite systems: $m_{u,d} \approx 2-4\,$MeV cf.\ $m_{p} \approx 1\,$GeV.  Evidently, the interaction energy greatly exceeds the rest masses of the anticipated constituents, making inapplicable the wave functions typical of Schr\"odinger quantum mechanics.

The difficulties appear chiefly because particle-number is not conserved by Lorentz boosts;
and extreme challenges are faced when constituents are light, \emph{e.g}.\ wave functions describing incoming and outgoing scattering states then represent systems with different particle content, so a probability interpretation is lost.  Such problems are circumvented by using a light-front formulation because eigenfunctions of the Hamiltonian are then independent of the system's four-momentum \cite{Keister:1991sb, Brodsky:1997de}.  

The light-front wave function of a hadron with momentum $P$ and spin $\lambda$, $\Psi(P,\lambda)$, is complicated.  In terms of perturbation theory's partons, $\Psi(P,\lambda)$ has a countably-infinite Fock-space expansion, with the $N$-parton term depending on $3N$ momentum variables, constrained such that their sum yields $P$, with a similar constraint on their spin (and flavour).  Were it necessary to use this complete object in analyses of even the simplest processes, then little connection between experiment and theory could be made.
Fortunately, collinear factorisation in the treatment of hard exclusive processes entails that much can be gained merely by studying hadron leading-twist parton distribution amplitudes (PDAs) \cite{Lepage:1979zb, Efremov:1979qk, Lepage:1980fj}.  Such a PDA is obtained from the simplest term in the Fock-space expansion, \emph{e.g}.\ meson, quark-antiquark ($\check N=2$) or baryon, three-quark ($\check N=3$), with the constituents' light-front-transverse momenta integrated out to a given scale, $\zeta$.


Regarding ground-state $S$-wave light-meson leading-twist PDAs, the last decade has seen real progress, not concerning their conformal limit  \cite{Lepage:1979zb, Efremov:1979qk, Lepage:1980fj}: $\varphi(x;\zeta) = 6 x(1-x)$, $m_{\rm p}/\zeta \simeq 0$; but on $m_p/\zeta \simeq 1$, where they are now known to be broad, concave functions, \emph{e.g}.\ $\varphi_\pi(x;\zeta\simeq m_p) \approx (8/\pi)\sqrt{x(1-x)}$ \cite{Brodsky:2006uqa, Chang:2013pq, Shi:2015esa, Braun:2015axa, Horn:2016rip, Gao:2016jka, Zhang:2017bzy, Mezrag:2016hnp}.  This resolves a long-time conflict, eliminating the notion that such PDAs exhibit a minimum at zero relative momentum \cite{Chernyak:1983ej}.  

Concerning the proton's leading-twist PDA, however, the situation is as unsatisfactory today as it was for mesons ten years ago.  Estimates of low-order Mellin moments exist, obtained using sum rules \cite{Chernyak:1983ej, Stefanis:1992nw} or lattice-QCD (lQCD) \cite{Braun:2008ur, Braun:2014wpa, Bali:2015ykx}, but there are no quantum field theory computations of this PDA's pointwise behaviour; and nothing is known about the PDA of the proton's radial excitation.  These issues are addressed herein.

\smallskip

\noindent\emph{2.\,Proton PDA: Definition}\,---\,%
In the isospin-symmetry limit, the proton possesses one independent leading-twist (twist-three) PDA \cite{Braun:2000kw}, denoted $\varphi([x];\zeta)$ herein:
\begin{align}
\nonumber
& \langle 0 | \varepsilon^{abc}\, \tilde u_+^a(z_1) \, C^\dagger \slash\!\!\! n \, u_-^b(z_2) \,
    \slash\!\!\! n  \, d_+^c(z_3) | P, +\rangle \\
&    =: \tfrac{1}{2} f_p  \, n\cdot P  \,\slash\!\!\! n \, N_+ \!
    \int_0^1 [dx] \, \varphi([x];\zeta) {\rm e}^{-i n\cdot P \sum_i x_i z_i}
    \label{phidefn}
\end{align}
where
$n^2=0$;
$(a,b,c)$ are colour indices;
$\psi_\pm = H_\pm \psi := (1/2)(\mathbf{I}_{\rm D}\pm \gamma_5)  \psi$, $\slash\!\!\! n = \gamma\cdot n$;
$\tilde q$ indicates matrix transpose;
$C$ is the charge conjugation matrix, $N=N(P)$ is the proton's Euclidean Dirac spinor (Ref.\,\cite{Segovia:2014aza}, Appendix\,B);
$\int_0^1 [dx] f([x])= \int _0^1 dx_1 dx_2 dx_3 \delta(1-\sum_i x_i)f([x])$;
and $f_p$ 
measures the proton's ``wave function at the origin''.

$\varphi([x])$ can be computed once the proton's Poincar\'e-covariant wave function is in hand; and following thirty years of study \cite{Cahill:1988dx, Burden:1988dt, Cahill:1988zi, Reinhardt:1989rw, Efimov:1990uz}, a clear picture has appeared.  At an hadronic scale, the proton is a Borromean system, bound by two effects \cite{Segovia:2015ufa}: one originates in non-Abelian facets of QCD, expressed in the effective charge \cite{Binosi:2016nme} and generating confined, nonpointlike but strongly-correlated colour-antitriplet diquarks in both the isoscalar-scalar and isotriplet-pseudovector channels; and that attraction is magnified by quark exchange associated with diquark breakup and reformation.  The presence and character of the diquarks owe to the mechanism that dynamically breaks chiral symmetry in the Standard Model \cite{Segovia:2015ufa}.

%
The proton Faddeev amplitude can be written \cite{Segovia:2014aza}:
\begin{equation}
\label{PsiP}
\Psi(P) = \psi_1 + \psi_2 + \psi_3\,,
\end{equation}
where the subscript identifies the bystander quark, \emph{i.e}.\ the quark not participating in a diquark, $\psi_3$ gives $\psi_{1,2}$ by cyclic permutation of all quark labels, and
\begin{subequations}
\label{psi3}
\begin{align}
\psi_3&(\{p\},\{\alpha\},\{\sigma\})  = {\mathpzc N}_{\,3}^{0} + {\mathpzc N}_{\,3}^{1}, \\
{\mathpzc N}_{\,3}^{0} & =
\big[ \Gamma^{0}(k;K) \big]_{\sigma_1 \sigma_2}^{\alpha_1\alpha_2}
\Delta^{0}(K) \big[ {\mathpzc S}(\ell;P) u(P) \big]_{\sigma_3}^{\alpha_3},\\
{\mathpzc N}_{\,3}^{1} & =
\big[ \Gamma_\mu^{1j}(k;K) \big]_{\sigma_1 \sigma_2}^{\alpha_1\alpha_2}
\Delta_{\mu\nu}^{1}(K) \big[ {\mathpzc A}_\nu^j(\ell;P) u(P) \big]_{\sigma_3}^{\alpha_3},
\end{align}
\end{subequations}
$(\{p\},\{\alpha\},\{\sigma\}) $ are the momentum, isospin and spin labels of the dressed-quarks constituting the bound state;
$P=p_1 + p_2 + p_3$ is the total momentum of the baryon;
$k=p_1$, 
$K=p_1+p_2$, $\ell = -K + (2/3) P$;
and the $j$ sum runs over the $(1,1)=+$ and $(1,0)=0$ isospin projections.
The matrix-valued functions $\Gamma$ in Eqs.\,\eqref{psi3} are diquark correlation amplitudes;
$\Delta^{0}$, $\Delta_{\mu\nu}^{1}$ are associated dressed-propagators;
and ${\mathpzc S}$, ${\mathpzc A}_\mu^j$ are matrix-valued quark-diquark amplitudes, describing the relative-momentum correlation between the diquark and bystander quark.

The proton's Faddeev wave function, $\chi$, is obtained from Eqs.\,\eqref{PsiP}, \eqref{psi3} by attaching the appropriate dressed-quark and -diquark propagators.  Each quantity involved is known because the nucleon Faddeev equation has been widely studied \cite{Segovia:2014aza, Xu:2015kta, Segovia:2015hra, Segovia:2016zyc, Eichmann:2016hgl, Lu:2017cln, Chen:2017pse}.  We therefore proceed by using algebraic representations for every element, with each form and their relative strengths, when combined, based on these analyses.
The dressed-quark propagator $S(p)=(-i\slash\!\!\! p + M)\sigma_{M}(p^2)$, $\sigma_M(s) = 1/[s+M^2]$, $\hat\sigma_M(s) = M^2/[s+M^2]$; $\Delta^{0}(K)=\sigma_{M_0}(K^2)$, $\Delta_{\mu\nu}^{1}(K)=T_{\mu\nu}(K)\sigma_{M_1}(K^2)$, $T_{\mu\nu}(K)=[\delta_{\mu\nu}+K_\mu K_\nu/K^2]$; 
{\allowdisplaybreaks
\begin{subequations}
\label{G0G1defn}
\begin{align}
{\mathpzc n}_0 \Gamma^0(k;K)C^\dagger & = i  \gamma_5 \! \int_{-1}^1dz\,\rho(z)\,\hat\sigma_{\Lambda_\Gamma}(k_{+K}^2)\,,\\
{\mathpzc n}_1 \Gamma_\mu^1(k;K)C^\dagger & = i (\gamma_\mu^{\rm T}
    + {\mathpzc r}_1 {\mathpzc f}(k;K) [ \slash\!\!\! k,\gamma_\mu^{\rm T}] ) \nonumber \\
    & \quad \times  \!  \int_{-1}^1dz\,\rho(z)\,\hat\sigma_{\Lambda_\Gamma}(k_{+K}^2)\,,
\end{align}
\end{subequations}
}
\hspace*{-0.5\parindent}where
$\rho(z) = (3/4) (1-z^2)$, $k_{+K} = k + (z-1) K/2$;
$\gamma_\mu^{\rm T} =T_{\mu\nu}(K) \gamma_\nu$, ${\mathpzc f}(k;K)= k\cdot K/(k^2 K^2 (k-K)^2)^{1/2}$;
and ${\mathpzc r}_1=1/4$, ${\mathpzc n}_{0,1}$ are fixed by requiring that the zeroth Mellin moment of the leading-twist PDA of each diquark correlation is $[n\cdot K/n\cdot P]$.
The final elements are:
\begin{subequations}
\label{SAdefn}
\begin{align}
{\mathpzc n} \, {\mathpzc S}(\ell;P) & = i  \! \int_{-1}^1dz\,\rho(z)\,\hat\sigma_{\Lambda_p^0}(w_{+P})\,,\\
\nonumber
{\mathpzc n} \, {\mathpzc A}_\nu^j(\ell;P) &= {\mathpzc r}_{\mathpzc A}
\tfrac{1}{6}{\mathpzc o}^j \gamma_5 [ \gamma_\nu - i {\mathpzc r}_{P} P_\nu] \\
& \quad \times  \int_{-1}^1dz\,\rho(z)\,\hat\sigma_{\Lambda_p^1}(w_{+P})\,,
\end{align}
\end{subequations}
where
$w_{+P} = [-\ell_{+P} + (2/3) P]^2$;
${\mathpzc o}^+  = \surd 2$, ${\mathpzc o}^0 = -1$;
${\mathpzc r}_{P} = 13/87$;
${\mathpzc r}_{\mathpzc A}$ measures the relative $1^+$:$0^+$ diquark strengths in the Faddeev amplitude;
and ${\mathpzc n}$ is that amplitude's canonical normalisation constant, whose value ensures the proton has unit charge \cite{Oettel:1999gc}.

Eqs.\,\eqref{G0G1defn}, \eqref{SAdefn} define a constrained spectral function model for $\chi$ \cite{Segovia:2014aza, Xu:2015kta, Segovia:2015hra, Segovia:2016zyc, Eichmann:2016hgl, Lu:2017cln, Chen:2017pse}, whose fidelity will subsequently be tested.  Crucially, the form is completely general: one can always use perturbation theory integral representations (PTIRs) for the propagators and amplitudes that arise in solving the continuum bound-state problem \cite{Nakanishi:1963zz, Nakanishi:1969ph, Nakanishi:1971}; so our subsequent analysis will establish an archetype for the continuum computation of baryon PDAs.

\begin{figure*}[!t]
\begin{center}
\begin{tabular}{lcrcrcr}
\parbox[c]{0.22\linewidth}{\includegraphics[clip,width=\linewidth]{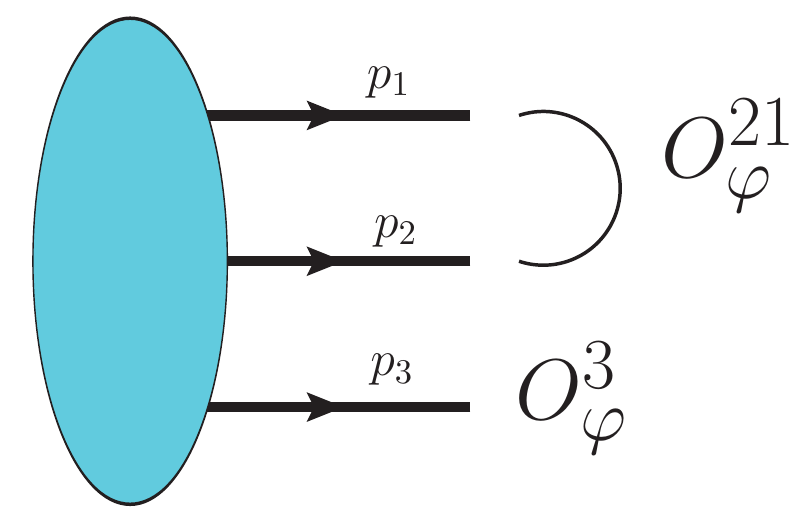}} & $\propto$\hspace*{0em} &
\parbox[c]{0.22\linewidth}{\includegraphics[clip,width=\linewidth]{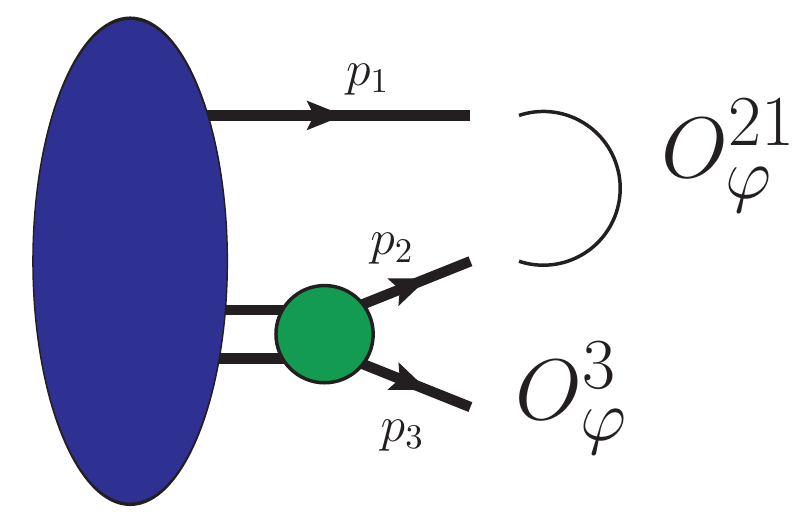}}\vspace*{-0ex}& $+$\hspace*{0em} &
\parbox[c]{0.22\linewidth}{\includegraphics[clip,width=\linewidth]{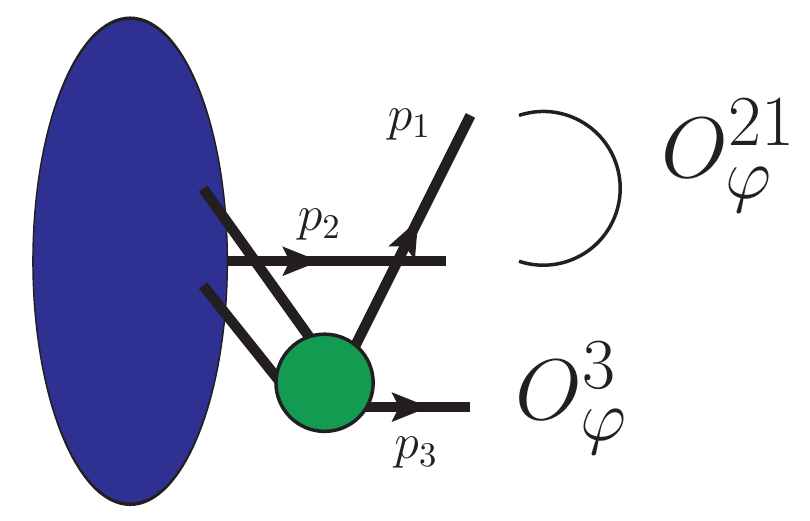}} & $+$\hspace*{0em} &
\parbox[c]{0.22\linewidth}{\includegraphics[clip,width=\linewidth]{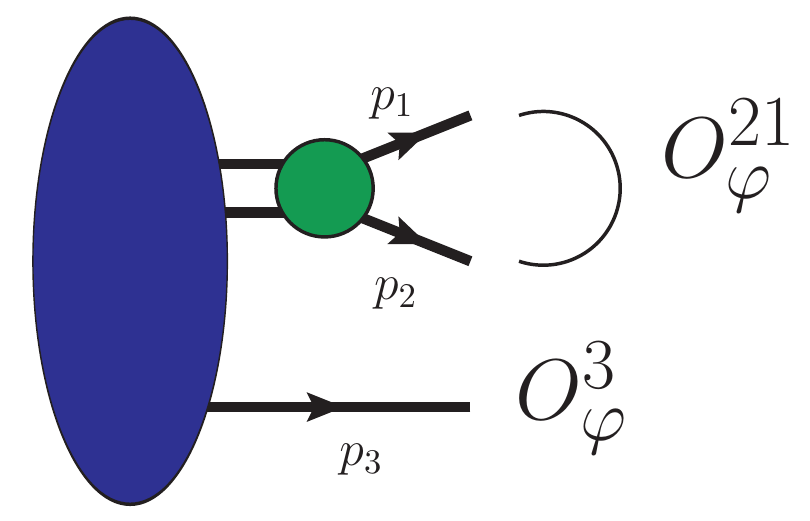}}
\end{tabular}
\end{center}
\caption{\label{PDAimage}
Studies of the continuum three-body bound-state problem reveal that diquark correlations are an integral part of the proton's Poincar\'e-covariant wave function, Eqs.\,\eqref{PsiP}, \eqref{psi3}, in which case Eq.\,\eqref{phidefn} is the sum of three terms, with the spinor projection operators given in Eq.\,\eqref{Odefn}.  The large (dark blue) ovals represent the $({\mathpzc S,\mathpzc A})$ elements in $\psi_{1,2,3}$, the (green) circles are the diquark correlation amplitudes, and the single and double lines are dressed-quark and -diquark propagators, respectively.
}
\end{figure*}

\smallskip

\noindent\emph{3.\,Proton PDA: Calculation}\,---\,%
Whenever the proton's Faddeev amplitude is as specified by Eqs.\,\eqref{PsiP}, \eqref{psi3}, then Eq.\,\eqref{phidefn} can be written as depicted in Fig.\,\ref{PDAimage}, where
\begin{equation}
\label{Odefn}
{\mathpzc O}_\varphi^{21} = H_-\,C^\dagger\,\slash\!\!\! n \, H_+\,,\;
{\mathpzc O}_\varphi^{3} = \slash\!\!\! n \, H_+\,.
\end{equation}
As a concrete illustration, we consider the first diagram on the rhs, whose contribution to the proton's PDA is fully determined by the following Mellin moments:
\begin{align}
\nonumber
& \tfrac{f_p}{2} \, n\cdot P \, \slash\!\!\! n \, N_+ \int [dx] \, x_1^l x_2^m\, \varphi([x])  
=: \tfrac{f_p}{2} \, n\cdot P\,  \slash\!\!\! n  \, N_+ \langle x_1^l x_2^m \rangle \\
\nonumber
& =  \int [dx] \, x_1^l x_2^m\,  \int\frac{d^4\ell}{(2\pi)^4}\frac{d^4k}{(2\pi)^4}
\delta_n^{x_1}(\ell + P/3) \delta_n^{x_2}(k) \\
&
\quad \quad \times \chi_1(\{p\},\{\alpha\},\{\sigma\})  \, {\mathpzc O}_\varphi^{21} \, {\mathpzc O}_\varphi^{3}\,,
\label{scalar1}
\end{align}
where $\delta_n^{x}(p) = \delta(n\cdot p- x n\cdot P)$; $p_1=\ell + P/3$, $p_2=k$, $p_3=K-k$.  Considering only the $0^+$ diquark component, the second and third contributions in Fig.\,\ref{PDAimage} vanish because this correlation is isoscalar-scalar, and hence the leading-twist part of the last line in Eq.\,\eqref{scalar1} is $\gamma\cdot {\mathpzc L}^{0^+}$, \begin{align}
\nonumber
& {\mathpzc L}_{\mu}^{0^+}  =
\tfrac{1}{4} {\rm tr}_{\rm D}\left[S_d(p_3) \Gamma^0(k;K) \tilde S_u(p_2) H_- C^\dagger \slash\!\!\! n \, H_+ \right. \\
& \quad\quad\quad  \times \left. S_u(p_1)\Delta^0(K){\mathpzc S}(\ell;P) \gamma_\mu \slash\!\!\! n \, H_+  \right]\\
\nonumber
& = \tfrac{1}{4} {\rm tr}_{\rm D}\left[ \gamma_\mu \slash\!\!\! n \, H_+ S_u(p_1)\Delta^0(K){\mathpzc S}(\ell;P)\right] \\
& 
\quad \times \tfrac{1}{2} {\rm tr}_{\rm D}\left[S_d(p_3) \Gamma^0(k;K) \tilde S_u(p_2) C^\dagger \gamma_5 \slash\!\!\! n \right],
\label{diquarkPDA}
\end{align}
where we have used properties of ${\rm tr}_{\rm D}$, the projection operators $H_\pm$, and $n_\mu$.  Inserting Eq.\,\eqref{diquarkPDA} into Eq.\,\eqref{scalar1}, one finds that the scalar diquark contribution to the proton's PDA is obtained from a convolution of the diquark's PDA with that of the bystander-quark in the quark+diquark Faddeev amplitude.  Importantly, the result generalises to the isotriplet-pseudovector component of the proton's Faddeev wave function, in which case the proton's PDA receives contributions from all three diagrams in Fig.\,\ref{PDAimage}.

Continuing our illustrative calculation, one first computes the scalar diquark PDA following the methods described in Refs.\,\cite{Chang:2013pq, Mezrag:2016hnp}.  Namely, in the $k$-integration of Eq.\,\eqref{scalar1}, use a Feynman parametrisation to rearrange the integrand such that there is a single denominator, a $k$-quadratic form raised to some power; and employ a suitably chosen change of variables in order to evaluate the integral over this relative four-momentum using standard algebraic methods.  This yields, with $\bar u = 1-u$ and $z=-1+2 [\bar u -\beta]/[\bar u - v]$,
\begin{subequations}
\label{D0mdefn}
\begin{align}
\nonumber
& {\mathpzc D}_0^m(K^2)  = {\mathpzc n}_0^\prime (K^2) \left[\frac{n\cdot K}{n\cdot P}\right]^{1+m}\\
& \times \int
\frac{du \, dv \, d\beta \,  \beta^m \, \rho(z(u,v,\beta)) 2 M}
{[\beta(v[\beta-2]+\beta)+\bar u (v-\beta^2)][K^2+{\mathpzc M}^2]}\,,\\
& {\mathpzc M}^2  =
\frac{[1-\bar u + v] M^2 + [\bar u - v]\Lambda_\Gamma^2}
{\beta(v[\beta-2]+\beta)+\bar u (v-\beta^2)}[\bar u - v]\,.
\end{align}
\end{subequations}

In our case, one can straightforwardly obtain the following algebraic result when $\Lambda_\Gamma = M$ ($\hat x_3=1-\hat x_2$, $\hat x_2=x_2/[x_2+x_3]$, $y=M^2/K^2$):
\begin{equation}
\label{PDAqq}
{\mathpzc n}_0^{\prime\prime}(K^2)
\varphi_{0}(\hat x_2,\hat x_3) = 12 y  ( 1 - \frac{y}{\hat x_2 \hat x_3 } \ln [ 1 + \hat x_2 \hat x_3/y ]) \,,
\end{equation}
where ${\mathpzc n}_0^{\prime\prime}(K^2)$ ensures $1=\int dx  \varphi_{0}(x,1-x)$ at each $K^2$.  Notably, when $K^2\ll \Lambda_\Gamma^2$, $\varphi_{0}(\hat x_2,\hat x_3) = 6 \hat x_2 \hat x_3$, \emph{viz}.\ the two-body conformal-limit PDA, which describes a correlation-free system; whereas on $K^2\gg \Lambda_\Gamma^2$, $\varphi_{0}(\hat x_2, \hat x_3) = 1$, which is the PDA of a pointlike two-body composite, the most highly-correlated system possible.

Using Eqs.\,\eqref{D0mdefn}, suppressing ${\mathpzc n}$ in Eq.\,\eqref{SAdefn}, one can rewrite Eq.\,\eqref{scalar1} in the form ($p_1=\ell + P/3$, $K=-p_1+P$):
\begin{align}
\nonumber
  \tfrac{f_p}{2} \, [n\cdot P]^2 &  \slash\!\!\! n \,N_+ \langle x_1^l x_2^m \rangle =  \int\frac{d^4\ell}{(2\pi)^4} \left[\frac{n\cdot p_1}{n\cdot P}\right]^l \\
%
&
 \times \slash\!\!\! n  \, H_+ S_u(p_1)
\Delta^0(K) {\mathpzc S}(\ell;P) {\mathpzc D}_0^m(K^2)
\label{scalar2} \,,
\end{align}
at which point the analysis leading to Eqs.\,\eqref{D0mdefn} can be adapted to solve this final ``two-body'' (quark+diquark) convolution problem for the $0^+$-diquark component of $\varphi([x];\zeta)$.  The result is an equation that expresses this contribution to $\varphi([x];\zeta)$ as an integral over five Feynman parameters in which the denominator is a single $\ell$-quadratic form.
%
%
The complete result for $\varphi([x];\zeta)$ is obtained by adding the $1^+$-diquark contributions.  That is readily accomplished by employing the procedure sketched above.   The addition is a sum of three integrals, two involving seven Feynman parameters, the third, nine, and each with a denominator that is an $\ell$-quadratic form.

\begin{figure*}[!t]
\begin{center}
\begin{tabular}{ccc}
\parbox[c]{0.31\linewidth}{\includegraphics[clip,width=\linewidth]{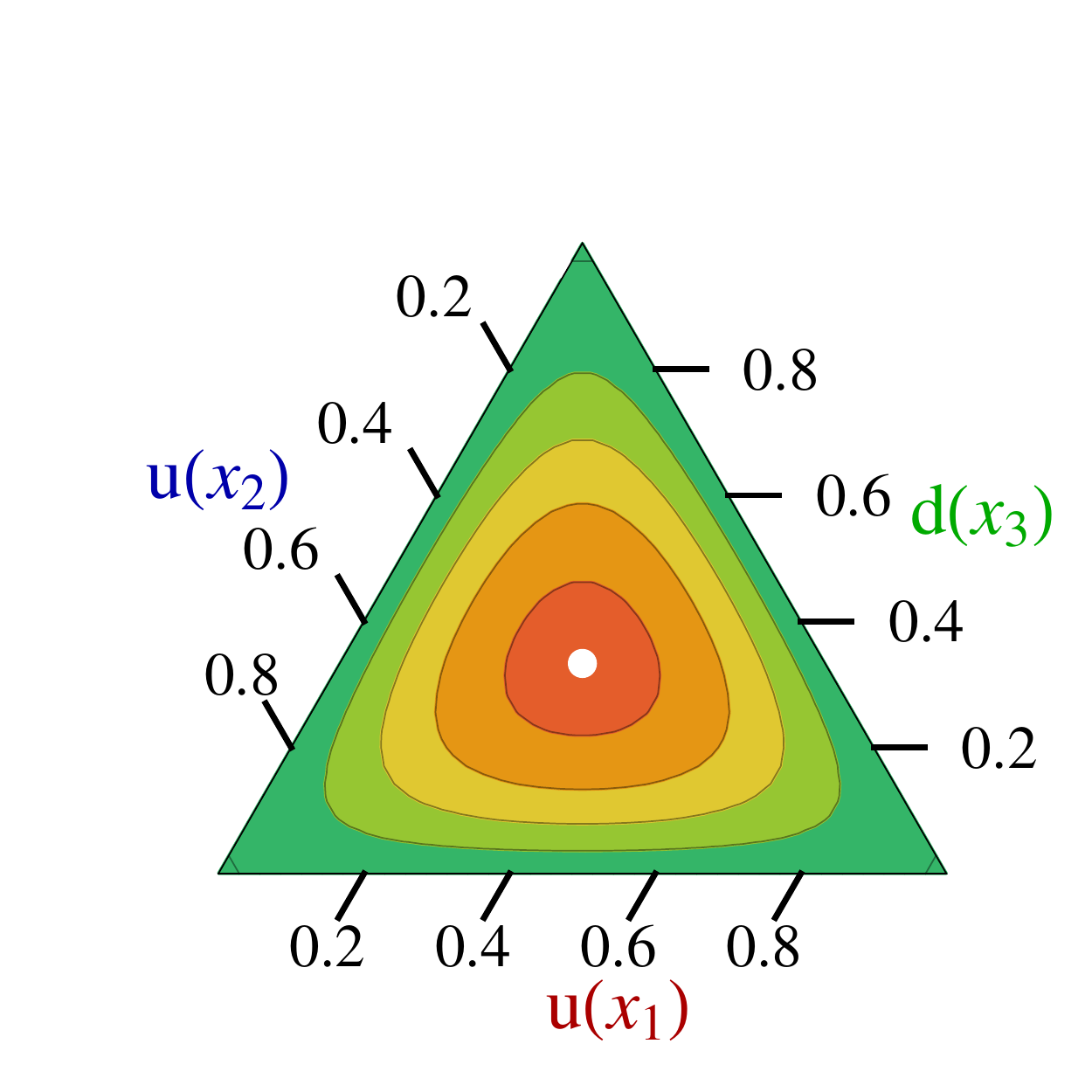}} &
\parbox[c]{0.31\linewidth}{\includegraphics[clip,width=\linewidth]{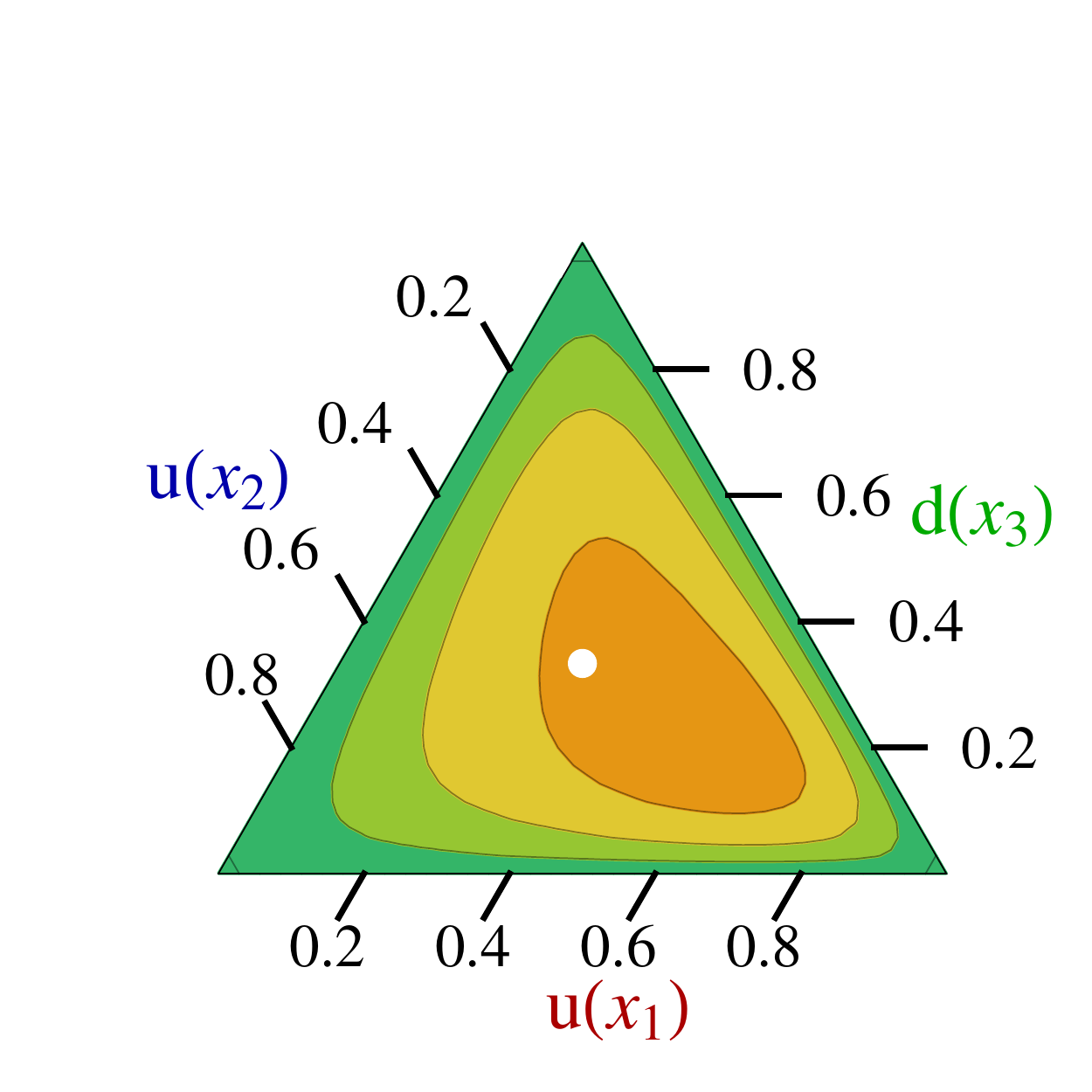}} &
\parbox[c]{0.37\linewidth}{\includegraphics[clip,width=\linewidth]{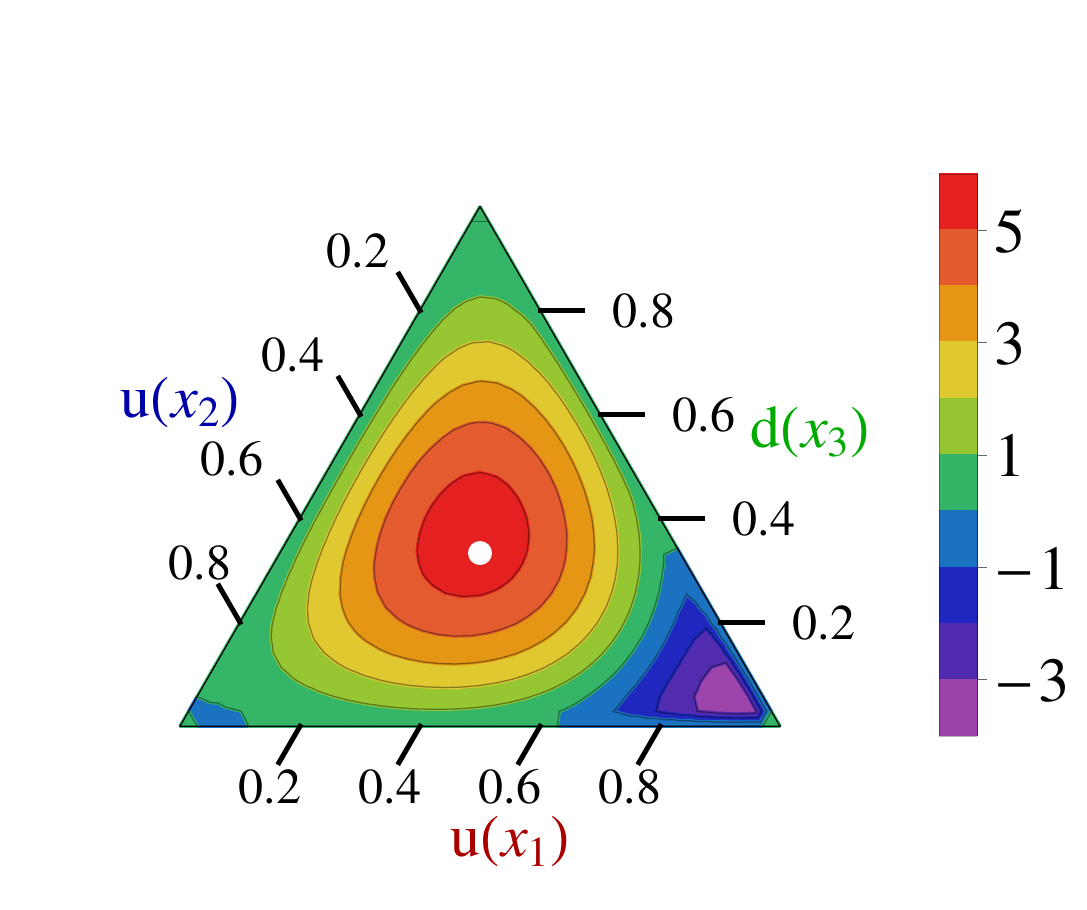}}
\end{tabular}\vspace*{-5ex}
\end{center}
\caption{\label{PlotPDAs} Barycentric plots:
\emph{left panel} -- conformal limit PDA, $\varphi_N^{\rm cl}([x])=120 x_1 x_2 x_3$; \emph{middle panel} -- computed proton PDA evolved to $\zeta=2\,$GeV, which peaks at $([x])=(0.55,0.23,0.22)$; and \emph{right panel} -- Roper resonance PDA at $\zeta=2\,$GeV.  The white circle in each panel serves only to mark the centre of mass for the conformal PDA, whose peak lies at $([x])=(1/3,1/3,1/3)$.
}
\end{figure*}

 All integrals required to compute $\varphi([x];\zeta)$ are readily evaluated numerically.  We choose the parameters in Eqs.\,\eqref{G0G1defn}, \eqref{SAdefn} so as to emulate realistic Faddeev wave functions \cite{Segovia:2014aza, Xu:2015kta, Segovia:2015hra, Segovia:2016zyc, Eichmann:2016hgl, Lu:2017cln, Chen:2017pse}:
$M=2/5$, $M_0=2/3$, $M_1=3/4$, $\Lambda_\Gamma=2/5$, $\Lambda_p^0=1$, $\Lambda_p^1=2/5$,
 in units of $m_p$, with ${\mathpzc r}_{\mathpzc A} = 0.30\pm 0.03$ ensuring that the scalar diquark contribution to the proton's baryon number is 65$\,\pm\,$5\%.  The distribution thus obtained is that associated with the hadronic scale $\zeta=\zeta_H = 0.51\,$GeV \cite{Chen:2016sno}.

We evolve $\varphi([x];\zeta_H)$ to $\zeta=\zeta_2=2\,$GeV by adapting the algorithm in Refs.\,\cite{Chang:2013pq, Shi:2015esa} to the case of baryons, \emph{i.e}.\ generalising the functional representation in Ref.\,\cite{Braun:1999te} and using the leading-order evolution equation in Ref.\,\cite{Lepage:1980fj}.  The result is depicted in Fig.\,\ref{PlotPDAs} and efficiently interpolated using ($w_{00}=1$)
{\allowdisplaybreaks
\begin{eqnarray}
\nonumber
\varphi([x]) & = & {\mathpzc n}_{\varphi} \, x_1^{\alpha_-} (x_2 x_3)^{\beta_-}
\sum_{j=0}^2\sum_{i=0}^j w_{ij} \, P_{j-i}^{2[i+\beta];\alpha_-}(2x_1-1) \\
&& \times  (x_2+x_3)^i C_i^\beta([x_3-x_2]/[x_2+x_3])\,, \label{EqInterpolation}
\end{eqnarray}
}
\hspace*{-0.5\parindent}where
${\mathpzc n}_{\varphi}$ ensures $\int [dx] \varphi([x])=1$;
$(\alpha,\beta)_- = (\alpha,\beta)-1/2$;
$P$ is a Jacobi function, $C$ a Gegenbauer polynomial;
and the interpolation parameters are listed in Table\,\ref{interpolation}A.

\begin{table}[!b]
\caption{\emph{A} --  Eq.\,\eqref{EqInterpolation} interpolation parameters for the proton and Roper PDAs in Fig.\,\ref{PlotPDAs}.  
%
\emph{B} -- Computed values of the first four moments of the PDAs.  Our error on $f_N$ reflects a scalar diquark content of $65\pm 5$\%; and values in rows marked with ``$\not\supset \mbox{av}$'' were obtained assuming the baryon is constituted solely from a scalar diquark.
(All results listed at $\zeta=2\,$GeV.)
\label{interpolation}
}
\begin{tabular*}
{\hsize}
{
l|@{\extracolsep{0ptplus1fil}}
c|@{\extracolsep{0ptplus1fil}}
c|@{\extracolsep{0ptplus1fil}}
c|@{\extracolsep{0ptplus1fil}}
c|@{\extracolsep{0ptplus1fil}}
c|@{\extracolsep{0ptplus1fil}}
c|@{\extracolsep{0ptplus1fil}}
c|@{\extracolsep{0ptplus1fil}}
c@{\extracolsep{0ptplus1fil}}}\hline
A & ${\mathpzc n}_{\hat\varphi}$ & $\alpha$ & $\beta$ & $w_{01}$ & $w_{11}$ & $w_{02}$ & $w_{12}$ & $w_{22}$ \\\hline
$p$ & 65.8 & 1.47 & 1.28 & $\phantom{-}0.096$ & $0.094$ & $\phantom{-}0.15$ & $-0.053$ & $\phantom{-}0.11$ \\
$R$ & 14.4 & 1.42 & 0.78 & $-0.93\phantom{6}$ & $0.22\phantom{0}$ & $-0.21$ & $-0.057$ & $-1.24$ \\\hline
\end{tabular*}
\smallskip

\begin{tabular*}
{\hsize}
{
l|@{\extracolsep{0ptplus1fil}}
c|@{\extracolsep{0ptplus1fil}}
c|@{\extracolsep{0ptplus1fil}}
c|@{\extracolsep{0ptplus1fil}}
c@{\extracolsep{0ptplus1fil}}}\hline
B & $10^3 f_N/\mbox{GeV}^2$ & $\langle x_1\rangle_u$ & $\langle x_2\rangle_u$ & $\langle x_3\rangle_d$ \\\hline
conformal PDA & & $0.333\phantom{(99)}$ & $0.333\phantom{(9)}$ & $0.333\phantom{(9)}$ \\\hline
lQCD \mbox{\cite{Braun:2014wpa}} & $2.84(33)$ & $0.372(7)\phantom{9}$ & 0.314(3) & 0.314(7) \\
lQCD \mbox{\cite{Bali:2015ykx}} & $3.60(6)\phantom{9}$ &  $0.358(6)\phantom{9}$ & $0.319(4)$ & 0.323(6) \\\hline
herein proton & $3.78(14)$ & $0.379(4)\phantom{9}$ & 0.302(1) & 0.319(3) \\
herein proton $\not\supset \mbox{av}$ & $2.97\phantom{(17)}$ & $0.412\phantom{(17)}$ & $0.295\phantom{(7)}$ & $0.293\phantom{(7)}$ \\\hline\hline
herein Roper  & $5.17(32)$ & $0.245(13)$ & $0.363(6)$ & $0.392(6)$ \\
herein Roper $\not\supset \mbox{av}$ & $2.63\phantom{(14)}$ & $0.010\phantom{(19)}$ & $0.490\phantom{(9)}$ & $0.500\phantom{(9)}$ \\\hline
\end{tabular*}
\end{table}

Table\,\ref{interpolation}B lists the four lowest-order moments of our proton PDA.  They reveal valuable insights, \emph{e.g}.\ when the proton is drawn as solely a quark+scalar-diquark correlation, $\langle x_2 \rangle_u=\langle x_3 \rangle_d$, because these are the two participants of the scalar quark+quark correlation; and the system is very skewed, with the PDA's peak being shifted markedly in favour of $\langle x_1 \rangle_u > \langle x_2 \rangle_u$.  This outcome conflicts with lQCD results \cite{Braun:2014wpa, Bali:2015ykx}.  On the other hand, realistic Faddeev equation calculations indicate that pseudovector diquark correlations are an essential part of the proton's wave function.  Naturally, when these $\{uu\}$ and $\{ud\}$ correlations are included, momentum is shared more evenly, shifting from the bystander $u(x_1)$ quark into $u(x_2)$, $d(x_3)$.  Adding these correlations with the known weighting, the PDA's peak moves back toward the centre and our computed values of the first moments align with those obtained using lQCD.  This confluence delivers a significantly more complete understanding of the lQCD simulations, which are now seen to validate a picture of the proton as a bound-state with both strong scalar \emph{and} pseudovector diquark correlations, in which the scalar diquarks are responsible for $\approx 60$\% of the Faddeev amplitude's canonical normalisation.  Importantly, as found with ground-state $S$-wave mesons \cite{Chang:2013pq, Shi:2015esa, Braun:2015axa, Horn:2016rip, Gao:2016jka, Zhang:2017bzy}, the leading-twist PDA of the ground-state nucleon is both broader than $\varphi_N^{\rm cl}([x])$ and decreases monotonically away from its maximum in all directions.

Our framework is readily extended to describe the quark core of the proton's first radial excitation: $m_R=(3/2) m_p$ \cite{Segovia:2015hra, Chen:2017pse}.  The scalar functions in this system's Faddeev amplitude possess a zero at quark-diquark relative momentum $\surd \ell^2 \approx 0.4\,$GeV$\approx 1/[0.5\,{\rm fm}]$.  This feature can be implemented via Eq.\,\eqref{SAdefn}: $\rho(z) = (1-z^2) \to \rho_R(z) = {\mathpzc s}_R^{qq} (1-z^2) (z+z_R^{qq})$, where $({\mathpzc s}_R^{\mathpzc S},z_R^{\mathpzc S})=(-1,1/50)=({\mathpzc s}_R^{{\mathpzc A},\gamma},z_R^{{\mathpzc A},\gamma})$, $({\mathpzc s}_R^{{\mathpzc A},P},z_R^{{\mathpzc A},P})=(1,3/10)$ and $\Lambda_p^0 \to \Lambda_R^0 = 4/5$ were all fitted to reproduce known solutions for the first radial excitation.
We therewith obtain the PDA in the rightmost panel of Fig.\,\ref{PlotPDAs}, which is efficiently interpolated using Eq.\,\eqref{EqInterpolation} with the parameters in Table\,\ref{interpolation}A; and whose first four moments are listed in  Table\,\ref{interpolation}B.  This prediction reveals some curious features, \emph{e.g}.: the excitation's PDA is not positive definite and there is a prominent locus of zeros in the lower-right corner of the barycentric plot, both of which echo features of the wave function for the first radial excitation of a quantum mechanical system and have also been seen in the leading-twist PDAs of radially excited mesons \cite{Li:2016dzv, Li:2016mah}; and the impact of pseudovector correlations within this excitation is opposite to that in the ground-state, \emph{viz}.\ they shift momentum into $u(x_1)$ from $u(x_2)$, $d(x_3)$.

\smallskip


\noindent\emph{4.\,Epilogue}\,---\,%
%
We used simple perturbation theory integral representations (PTIRs) for all elements in the Faddeev wave functions, therewith defining models constrained by the best available solutions of the continuum three-valence-body bound-state equations.  Crucially, the technique we introduced is completely general: it can readily be used with any realistic Poincar\'e-covariant bound-state wave function, once it is expressed via PTIRs.  Hence, the veracity of our PDA predictions can straightforwardly be tested in future studies.
In the interim, the PDAs we have determined will, \emph{e.g}.\ enable the first realistic assessments to be made of the scale at which exclusive experiments involving baryons may properly be compared with predictions based on perturbative-QCD hard scattering formulae and thereby assist contemporary and planned facilities to refine and reach their full potential \cite{Dudek:2012vr, Burkert:2016dxc, Accardi:2012qut}.  The value of such estimates has recently been demonstrated in studies of mesons \cite{Chang:2013nia, Horn:2016rip, Gao:2017mmp}.

%

\smallskip

%
\noindent\emph{Acknowledgments}\,---\,We are grateful for insightful comments from V.\,Mokeev, H.\,Moutarde, F.\,Gao, S.-X.\,Qin, J.\,Rodr{\'i}guez-Quintero and S.-S.\,Xu.
Work supported by:
Argonne National Laboratory, Office of the Director's Postdoctoral Fellowship Program;
European Union's Horizon 2020 research and innovation programme under the Marie Sk\l{}odowska-Curie Grant Agreement No.\ 665919;
Spanish MINECO's Juan de la Cierva-Incorporaci\'on programme, Grant Agreement No. IJCI-2016-30028;
Spanish Ministerio de Econom\'ia, Industria y Competitividad under Contract Nos.\ FPA2014-55613-P and SEV-2016-0588;
the Chinese Government's Thousand Talents Plan for Young Professionals;
the Chinese Ministry of Education, under the \emph{International Distinguished Professor} programme;
and U.S.\ Department of Energy, Office of Science, Office of Nuclear Physics, under contract no.~DE-AC02-06CH11357;


\begin{thebibliography}{10}

\bibitem{Keister:1991sb}
B.~D. Keister and W.~N. Polyzou,
\newblock Adv. Nucl. Phys. {\bf 20}, 225 (1991).

\bibitem{Brodsky:1997de}
S.~J. Brodsky, H.-C. Pauli and S.~S. Pinsky,
\newblock Phys. Rept. {\bf 301}, 299 (1998).

\bibitem{Lepage:1979zb}
G.~P. Lepage and S.~J. Brodsky,
\newblock Phys. Lett. B {\bf 87}, 359 (1979).

\bibitem{Efremov:1979qk}
A.~V. Efremov and A.~V. Radyushkin,
\newblock Phys. Lett. B {\bf 94}, 245 (1980).

\bibitem{Lepage:1980fj}
G.~P. Lepage and S.~J. Brodsky,
\newblock Phys. Rev. D {\bf 22}, 2157 (1980).

\bibitem{Brodsky:2006uqa}
S.~J. Brodsky and G.~F. de~Teramond,
\newblock Phys. Rev. Lett. {\bf 96}, 201601 (2006).

\bibitem{Chang:2013pq}
L.~Chang {\em et~al.},
\newblock Phys. Rev. Lett. {\bf 110}, 132001 (2013).

\bibitem{Shi:2015esa}
C.~Shi {\em et~al.},
\newblock Phys. Rev. D {\bf 92}, 014035 (2015).

\bibitem{Braun:2015axa}
V.~M. Braun {\em et~al.},
\newblock Phys. Rev. D {\bf 92}, 014504 (2015).

\bibitem{Horn:2016rip}
T.~Horn and C.~D. Roberts,
\newblock J. Phys. G. {\bf 43}, 073001 (2016).

\bibitem{Gao:2016jka}
F.~Gao, L.~Chang and Y.-x. Liu,
\newblock Phys. Lett. B {\bf 770}, 551 (2017).

\bibitem{Zhang:2017bzy}
J.-H. Zhang, J.-W. Chen, X.~Ji, L.~Jin and H.-W. Lin,
\newblock Phys. Rev. D {\bf 95}, 094514 (2017).

\bibitem{Mezrag:2016hnp}
C.~Mezrag, H.~Moutarde and J.~Rodriguez-Quintero,
\newblock Few Body Syst. {\bf 57}, 729 (2016).

\bibitem{Chernyak:1983ej}
V.~L. Chernyak and A.~R. Zhitnitsky,
\newblock Phys. Rept. {\bf 112}, 173 (1984).

\bibitem{Stefanis:1992nw}
N.~G. Stefanis and M.~Bergmann,
\newblock Phys. Rev. D {\bf 47}, R3685 (1993).

\bibitem{Braun:2008ur}
V.~M. Braun {\em et~al.},
\newblock Phys. Rev. D {\bf 79}, 034504 (2009).

\bibitem{Braun:2014wpa}
V.~M. Braun {\em et~al.},
\newblock Phys. Rev. D {\bf 89}, 094511 (2014).

\bibitem{Bali:2015ykx}
G.~S. Bali {\em et~al.},
\newblock JHEP {\bf 02}, 070 (2016).

\bibitem{Braun:2000kw}
V.~Braun, R.~J. Fries, N.~Mahnke and E.~Stein,
\newblock Nucl. Phys. B {\bf 589}, 381 (2000),
\newblock [Erratum: Nucl. Phys. B\,\textbf{607}, 433 (2001)].

\bibitem{Segovia:2014aza}
J.~Segovia, I.~C. Clo{\"e}t, C.~D. Roberts and S.~M. Schmidt,
\newblock Few Body Syst. {\bf 55}, 1185 (2014).

\bibitem{Cahill:1988dx}
R.~T. Cahill, C.~D. Roberts and J.~Praschifka,
\newblock Austral. J. Phys. {\bf 42}, 129 (1989).

\bibitem{Burden:1988dt}
C.~J. Burden, R.~T. Cahill and J.~Praschifka,
\newblock Austral. J. Phys. {\bf 42}, 147 (1989).

\bibitem{Cahill:1988zi}
R.~T. Cahill,
\newblock Austral. J. Phys. {\bf 42}, 171 (1989).

\bibitem{Reinhardt:1989rw}
H.~Reinhardt,
\newblock Phys. Lett. B {\bf 244}, 316 (1990).

\bibitem{Efimov:1990uz}
G.~V. Efimov, M.~A. Ivanov and V.~E. Lyubovitskij,
\newblock Z. Phys. C {\bf 47}, 583 (1990).

\bibitem{Segovia:2015ufa}
J.~Segovia, C.~D. Roberts and S.~M. Schmidt,
\newblock Phys. Lett. B {\bf 750}, 100 (2015).

\bibitem{Binosi:2016nme}
D.~Binosi, C.~Mezrag, J.~Papavassiliou, C.~D. Roberts and
  J.~Rodriguez-Quintero,
\newblock Phys. Rev. D {\bf 96}, 054026 (2017).

\bibitem{Xu:2015kta}
S.-S. Xu {\em et~al.},
\newblock Phys. Rev. D {\bf 92}, 114034 (2015).

\bibitem{Segovia:2015hra}
J.~Segovia {\em et~al.},
\newblock Phys. Rev. Lett. {\bf 115}, 171801 (2015).

\bibitem{Segovia:2016zyc}
J.~Segovia and C.~D. Roberts,
\newblock Phys. Rev. C {\bf 94}, 042201(R) (2016).

\bibitem{Eichmann:2016hgl}
G.~Eichmann, C.~S. Fischer and H.~Sanchis-Alepuz,
\newblock Phys. Rev. D {\bf 94}, 094033 (2016).

\bibitem{Lu:2017cln}
Y.~Lu {\em et~al.},
\newblock Phys. Rev. C {\bf 96}, 015208 (2017).

\bibitem{Chen:2017pse}
C.~Chen {\em et~al.},
\newblock (2017),
\newblock {\emph{Structure of the nucleon's low-lying excitations},
  arXiv:1711.03142 [nucl-th]}.

\bibitem{Oettel:1999gc}
M.~Oettel, M.~Pichowsky and L.~von Smekal,
\newblock Eur. Phys. J. A {\bf 8}, 251 (2000).

\bibitem{Nakanishi:1963zz}
N.~Nakanishi,
\newblock Phys. Rev. {\bf 130}, 1230 (1963).

\bibitem{Nakanishi:1969ph}
N.~Nakanishi,
\newblock Prog. Theor. Phys. Suppl. {\bf 43}, 1 (1969).

\bibitem{Nakanishi:1971}
N.~Nakanishi,
\newblock {\em Graph Theory and Feynman Integrals} (Gordon and Breach, New
  York, 1971).

\bibitem{Chen:2016sno}
C.~Chen, L.~Chang, C.~D. Roberts, S.~Wan and H.-S. Zong,
\newblock Phys. Rev. D {\bf 93}, 074021 (2016).

\bibitem{Braun:1999te}
V.~M. Braun, S.~E. Derkachov, G.~P. Korchemsky and A.~N. Manashov,
\newblock Nucl. Phys. B {\bf 553}, 355 (1999).

\bibitem{Li:2016dzv}
B.~L. Li {\em et~al.},
\newblock Phys. Rev. D {\bf 93}, 114033 (2016).

\bibitem{Li:2016mah}
B.-L. Li, L.~Chang, M.~Ding, C.~D. Roberts and H.-S. Zong,
\newblock Phys. Rev. D {\bf 94}, 094014 (2016).

\bibitem{Dudek:2012vr}
J.~Dudek {\em et~al.},
\newblock Eur.\ Phys.\ J. A {\bf 48}, 187 (2012).

\bibitem{Burkert:2016dxc}
V.~D. Burkert,
\newblock Few Body Syst. {\bf 57}, 873 (2016).

\bibitem{Accardi:2012qut}
A.~Accardi {\em et~al.},
\newblock Eur. Phys. J. A {\bf 52}, 268 (2016).

\bibitem{Chang:2013nia}
L.~Chang, I.~C. Clo{\"e}t, C.~D. Roberts, S.~M. Schmidt and P.~C. Tandy,
\newblock Phys. Rev. Lett. {\bf 111}, 141802 (2013).

\bibitem{Gao:2017mmp}
F.~Gao, L.~Chang, Y.-X. Liu, C.~D. Roberts and P.~C. Tandy,
\newblock Phys. Rev. D {\bf 96}, 034024 (2017).

\end{thebibliography}

\end{document}